\documentclass[12pt]{article}
\usepackage{amsfonts}
\usepackage{amsmath}
\usepackage{amssymb}
\usepackage{graphicx}
\usepackage{color}
\usepackage[all, knot]{xy}
\usepackage{tikz}

\usepackage[utf8]{inputenc}
\usepackage{epstopdf}
\usepackage[footnotesize]{caption}
\usepackage{amsthm}
\usepackage{enumitem}
\usepackage{mathrsfs}

\usepackage[margin=3cm]{geometry}

\def \be {\begin{equation}}
\def \ee {\end{equation}}
\def \bea {\begin{eqnarray}}
\def \eea {\end{eqnarray}}
\def \nn {\nonumber}

\def \rr {\raise.35ex\hbox{\small $\prime$}\kern-.17em{\mbox{\large $\imath$}}}

\def \dels {\partial\kern-.6em /\kern.1em}
\def \As {{A\kern-.5em / \kern.5em}}
\def \Ds {D\kern-.7em / \kern.5em}

\def \ks {k\kern-.5em /}
\def \ls {l\kern-.5em /}

\newcommand{\ci}[1]{}

\newcommand{\ba}{\begin{eqnarray}}
\newcommand{\ea}{\end{eqnarray}}
\newcommand{\bal}{\begin{align}}
\newcommand{\eal}{\end{align}}
\newcommand{\bay}[1]{\left(\begin{array}{#1}}
\newcommand{\eay}{\end{array}\right)}

\newcommand{\hide}[1]{}

\newlist{axioms}{enumerate}{2}
\setlist[axioms,1]{label=\textbf{A\arabic{axiomsi}.}, ref=A\arabic{axiomsi}}
\setlist[axioms,2]{label=\textbf{A\arabic{axiomsi}\rlap{\myEnumCounter{axiomsii}}.},%
                   ref=A\arabic{axiomsi}\myEnumCounter{axiomsii},%
                   align=parleft,%
                   leftmargin=0em,%
                   itemsep=1.4ex,%
                   before={\stepcounter{axiomsi}}}
                   
  \usetikzlibrary{decorations.markings}

\begin{document}

\begin{titlepage}

\begin{center}

\hfill
\vskip .2in

\textbf{\LARGE
Lattice AdS Geometry and\\
 Continuum Limit
\vskip.3cm
}
\vskip .5in
{\large
Chen-Te Ma$^{a, b, c, d}$ \footnote{e-mail address: yefgst@gmail.com}
\\
\vskip 1mm
}
{\sl
$^a$
Guangdong Provincial Key Laboratory of Nuclear Science,\\ 
Institute of Quantum Matter, South China Normal University, Guangzhou 510006, Guangdong, China.
\\
School of Physics and Telecommunication Engineering,\\ 
South China Normal University, Guangzhou 510006, Guangdong, China.
\\
$^c$
The Laboratory for Quantum Gravity and Strings,\\
Department of Mathematics and Applied Mathematics,\\
University of Cape Town, Private Bag, Rondebosch 7700, South Africa.
\\
$^d$
Department of Physics and Center for Theoretical Sciences, \\
National Taiwan University, Taipei 10617, Taiwan, R.O.C..
}\\
\vskip 1mm
\vspace{40pt}
\end{center}
\newpage
\begin{abstract}
We construct the lattice AdS geometry. The lattice AdS$_2$ geometry and AdS$_3$ geometry can be extended from the lattice AdS$_2$ induced metric, which provided the lattice Schwarzian theory at the classical limit. Then we use the lattice embedding coordinates to rewrite the lattice AdS$_2$ geometry and AdS$_3$ geometry with the manifest isometry. The lattice AdS$_2$ geometry can be obtained from the lattice AdS$_3$ geometry through the compactification without the lattice artifact. The lattice embedding coordinates can also be used in the higher dimensional AdS geometry. Because the lattice Schwarzian theory does not suffer from the issue of the continuum limit, the lattice AdS$_2$ geometry can be obtained from the higher dimensional AdS geometry through the compactification, and the lattice AdS metric does not depend on the angular coordinates, we expect that the continuum limit should exist in the lattice Einstein gravity theory from this geometric lattice AdS geometry. Finally, we apply this lattice construction to construct the holographic tensor network without the issue of a continuum limit. 
\end{abstract}
\end{titlepage}

\section{Introduction}
\label{1}
\noindent 
Holographic principle \cite{Bousso:2002ju} states that the physical degrees of freedom of quantum gravity theory should be encoded on the boundary if phenomena related to the quantum gravity theory do not always occur inside the black hole \cite{tHooft:1993dmi}. The holographic principle was suggested by the area law of black hole entropy \cite{Strominger:1994tn}, which was found by generalizing the second law of ordinary thermodynamics \cite{Bekenstein:1972tm}, which states that entropy does not decrease in a closed system. The generalized second law of thermodynamics states that the sum of black hole entropy and environment entropy does not decrease in a closed system. The coefficient of black hole entropy was first computed by using the generalized second law of thermodynamics and assuming that the smallest possible radius of one particle is equal to its Compton wavelength \cite{Bekenstein:1973ur}. This computation did not provide the correct result. The correct understanding of black hole entropy after the four laws of black hole thermodynamics was obtained \cite{Bardeen:1973gs}, and the concepts of the Hawking temperature were proposed \cite{Hawking:1974sw}.\\

\noindent 
The holographic principle should be applied to the quantum gravity theory, not only to the classical gravity theory. Therefore, the evidence of the holographic principle from the black hole entropy should not be enough. The ultraviolet (UV) complete theory, string theory, showed that the area of a physical system can be related to the string coupling constant \cite{Susskind:1994vu}, which is defined by the dilaton field. This is the first evidence of the holographic principle from the UV complete theory. The string theory describes one-dimensional strings moving on the two-dimensional worldsheet. The fluctuation of target space provides the supergravity theory, which contains Einstein gravity theory, and provides a duality web to connect low-energy ten-dimensional theories. The concepts of duality web are also useful in the strongly correlated system \cite{Ma:2018efs}.\\

\noindent
 The solution of type IIB supergravity theory, five-dimensional anti-de Sitter spacetime (AdS$_5$) times the five-dimensional sphere, and its boundary theory, the four-dimensional $\mathcal{ N}$=4 U($N$) supersymmetric Yang-Mills theory \cite{Aharony:1999ti}, where $\mathcal{ N}$ is the number of supercharges \cite{Maldacena:1997re}, provides the conjecture, a theory in the $d$-dimensional anti-de Sitter spacetime is dual to the ($d-1$)-dimensional conformal field theory  (AdS$_d$/CFT$_{d-1}$ correspondence) \cite{Maldacena:1997re}. This conjecture also suggests that the infrared regulator of the bulk AdS theory corresponds to the ultraviolet regulator of the boundary theory, CFT$_{d-1}$ \cite{Susskind:1998dq}. Furthermore, the local symmetry in an AdS bulk theory corresponds to the global symmetry in CFT \cite{Witten:1998qj}. This was also found from the asymptotic symmetry of AdS$_3$ spacetime, which is the two-dimensional conformal group not SO(2, 2) \cite{Brown:1986nw}. This asymptotic symmetry was realized from the Poisson bracket algebra \cite{Brown:1986nw}.
\\

\noindent 
The recent concrete holographic study was found in the AdS$_2$ dilaton gravity theory \cite{Brown:1988am}, in which the theory is that a dilaton field is linearly coupled to Einstein gravity theory with the Dirichlet boundary condition and the negative cosmological constant \cite{Jackiw:1984je}. The boundary theory of the AdS$_2$ dilaton gravity theory, the Schwarzian term is linearly coupled to the dilaton field or the Schwarzian theory, which could be derived in the classical limit \cite{Jensen:2016pah}. The Schwarzian theory is also equivalent to the one-dimensional Liouville theory \cite{Engelsoy:2016xyb}. The major consistency in this holographic study is that the solution of the boundary dilaton field \cite{Cadoni:1994uf} could be obtained from the bulk and boundary theories consistently \cite{Maldacena:2016upp}. The Schwarzian theory is not the conformal field theory, but the AdS$_2$ dilaton gravity theory could be obtained from the AdS$_3$ Einstein gravity theory through the compactification \cite{Achucarro:1993fd}. Therefore, the holographic study in the AdS$_2$ dilaton gravity theory should provide strong evidence to the AdS/CFT correspondence.\\ 

\noindent 
The Schwarzian term can be built on the lattice. The lattice simulation of this lattice theory was already studied through the Metropolis \cite{Stanford:2017thb}, which is a numerical method and suffers from the ultralocal problem because one needs to smoothly transform a configuration in a site for each updating without too-low acceptance in the numerical study. The lattice theory cannot have the same symmetry as the isometry of the AdS$_2$ metric because the perturbation at the next non-trivial order of the lattice spacing breaks the symmetry \cite{Chu:2018jje}. Therefore, ones reconstructed a lattice theory with the complex fields and the same symmetry as the isometry of the AdS$_2$ metric, and the lattice perturbation could provide the consistent result to the action of discretized AdS$_2$ dilaton gravity theory up to the second-order of lattice spacing or boundary cut-off \cite{Chu:2018jje}. This lattice theory is called lattice Schwarzian theory \cite{Chu:2018jje}.\\

The non-trivial feature of lattice Schwarzian theory is that the manifest SL(2) invariant measure comes from the lattice AdS$_2$ induced metric \cite{Chu:2018jje}, which provides the AdS$_2$ induced metric at the non-trivial leading order with respect to the lattice spacing \cite{Jensen:2016pah}. The technical construction is similar to the lattice Liouville model \cite{Faddeev:2008xy}. The central question that we would like to address in this article is: {\it Could we construct the lattice AdS geometry with an exact isometry?} In this Letter, we construct the lattice AdS geometry without losing the isometry as in the lattice Schwarzian theory and define the lattice embedding coordinates to rewrite the lattice AdS geometry with the manifest isometry. This lattice embedding coordinates also provide the coordinate transformations to the lattice AdS geometry without losing the isometry. Finally, we also demonstrate that the lattice AdS$_2$ geometry can be exactly obtained from the lattice AdS$_3$ geometry through the compactification without any lattice artifact. The lattice Einstein gravity theory should not be renormalizable when dimensions of spacetime are larger than two so one cannot use the experiences from the renormalizable theory to understand. Our lattice construction provides the evidence to that the lattice Einstein gravity theory in the AdS background, which is a weakly coupled theory, should have the continuum limit, which is defined by the bare cut-off, at the infrared (IR) limit, which is defined by the physical cut-off, from the perturbative way.
\\

The direct application of the lattice geometry is the holographic entanglement entropy \cite{Ryu:2006bv}, which states that the universal term of entanglement entropy in the conformal field theory is given by the codimension two minimum surface at a given time slice in the bulk gravity theory. This conjecture of this minimum surface was shown from the geometry with the replica symmetry \cite{Holzhey:1994we} in the bulk gravity theory \cite{Lewkowycz:2013nqa}. This minimum surface motivates the tensor network \cite{Orus:2013kga} by doing the renormalization group flow in the real space through the identification between the minimum surface and the minimum number of bonds \cite{Swingle:2009bg}. The objects of the tensor network are tensors forming a network, which provides a quantum state. One example of the tensor network is the multi-scale entanglement renormalization ansatz (MERA), which applies the lattice coarse-graining transformations to remove the local entanglement in the real space. The MERA graph can be seen as a lattice AdS space at a given time slice. The most subtle issue is the continuum limit of lattice AdS geometry since the MERA graph is topological. This continuum limit meets a no-go theorem \cite{Bao:2015uaa}. In this Letter, we use the lattice AdS geometry to construct the tensor network without the continuum limit problem. Hence our study can be seen as the first principle of the holographic tensor network from the symmetry perspective.

\section{Lattice AdS Geometry}
\label{2}
We first construct the lattice AdS$_2$ geometry and AdS$_3$ geometry from the lattice AdS$_2$ induced metric \cite{Chu:2018jje}. Then we show that the lattice AdS geometry can be rewritten in terms of the lattice embedding coordinates without losing the isometry for each lattice spacing and size. We compactify the angular direction in the lattice AdS$_3$ geometry to obtain the lattice AdS$_2$ geometry without suffering from the lattice artifact. Hence the lattice AdS geometry without losing the isometry can be constructed from the lattice embedding coordinates without restricting to the lattice AdS$_2$ geometry and AdS$_3$ geometry. 

\subsection{Lattice AdS$_2$ Geometry and AdS$_3$ Geometry}
The lattice AdS$_2$ induced metric with the isometry for each lattice size and spacing \cite{Chu:2018jje} was given by 
\bea
h_{2\mathrm{l},j}=\frac{1}{\Lambda a_u^2}
\frac{(f_{j+1}^*-f^*_{j-1})(f_{j+1}-f_{j-1})}
{(f_{j+1}-f^*_{j+1})(f_{j-1}-f^*_{j-1})},
\eea
where $f_j\equiv t_j+iz_j$, $u_{j\pm 1}\equiv u_j\pm a_u$, $\Lambda<0$ is the cosmological constant, the coordinates $t$, $z$ are defined in the bulk AdS$_2$ geometry and are functions of the boundary time $u$, and the $a_u$ is the lattice spacing in the $u$ direction. The lattice sites in the $u$ direction are labeled by the index $j=1,\ 2,\ \cdots,\ n-1,\ n$, and the function $f$ satisfies the periodic boundary condition $f_0=f_{n+1}$. The number of lattice points is denoted by $n$. Because the AdS$_2$ spacetime interval is: $ds_2^2=\big(-1/\Lambda\big)(dt^2+dz^2)/z^2=\bigg(-1\big/\big(\Lambda z^2\big)\bigg)dx^{+}dx^-$, where $x^+\equiv t+iz$ and $x^-\equiv t-iz$, the lattice AdS$_2$ spacetime interval can be written as: 
\bea
&&\big(\Delta s_{2\mathrm{l},j,k,m_1, m_2, m_3, m_4}\big)^2
\nn\\
&\equiv& \frac{4}{\Lambda}\frac{g_{j+m_1,k+m_2}^*-g^*_{j+m_3,k+m_4}}{g_{j+m_1,k+m_2}-g^*_{j+m_1,k+m_2}}
\nn\\
&&\times \frac{g_{j+m_1,k+m_2}-g_{j+m_3,k+m_4}}{g_{j+m_3,k+m_4}-g^*_{j+m_3.k+m_4}}
\nn\\
&\equiv& 2g_{\mathrm{l},+-,j,k,m_1,m_2,m_3,m_4}
\nn\\
&&\times(g_{j+m_1,k+m_2}^*-g^*_{j+m_3,k+m_4})
\nn\\
&&\times(g_{j+m_1,k+m_2}-g_{j+m_3,k+m_4}),
\nn\\
\eea
where 
\bea
&&g_{j,k}\equiv t_j+iz_k,\qquad t_{j\pm 1}\equiv t_j\pm a_t,\qquad z_{k\pm 1}\equiv z_k\pm a_z, 
\nn\\
&& g_{\mathrm{l},+-,j,k,m_1,m_2,m_3,m_4}\equiv\frac{\frac{2}{\Lambda}}{(g_{j+m_1,k+m_2}-g^*_{j+m_1,k+m_2})(g_{j+m_3,k+m_4}-g^*_{j+m_3.k+m_4})}.
\nn\\
\eea
 The lattice AdS$_2$ spacetime interval has the similar form to the lattice AdS$_2$ induced metric so the lattice geometry still preserves the isometry for each lattice spacing and size as in the AdS$_2$ geometry. Because the lattice spacetime interval at each point is defined by two adjacent lattice points, the lattice spacetime interval depends on a choice of two adjacent lattice points. In other words, the lattice spacetime interval at each point depends on relations between the coordinate $t$ and the coordinate $z$. One can choose $m_1, m_2, m_3, m_4=1,\ 0, \ -1$.
\\

From the construction of the lattice AdS$_2$ geometry, we can quickly extend the construction to the lattice AdS$_3$ geometry without losing isometry for each lattice size and spacing. The lattice AdS$_3$ induced metric at a given time slice is defined by $h_{xx,3\mathrm{l},j}\equiv \bigg(1\big/\big(\Lambda a_x^2\big)\bigg)(f_{j+1}^*-f^*_{j-1})(f_{j+1}-f_{j-1})\big/\big((f_{j+1}-f^*_{j+1})(f_{j-1}-f^*_{j-1})\big)$, where $f_j\equiv x_j+iz_j$ and $x_{j\pm 1}\equiv x_j\pm a_x$, when the coordinate $z$ is the function of the coordinate $x$. The function $f_j$ does not necessary to satisfy the periodic boundary function. If one wants to consider that the coordinate $x$ is a function of the coordinate $z$, one only needs to replace $x_{j\pm 1}= x_j\pm a_x$ with $z_{j\pm 1}\equiv z_j\pm a_z$  and also replace $a_x^2$ with $a_z^2$ in the lattice AdS$_3$ induced metric at a given time slice. The AdS$_3$ metric is: $ds_3^2=-\big(1/\Lambda\big)\big(dt^2+dx^2+dz^2\big)/z^2=-\big(1/(\Lambda z^2)\big)(dt^2+dy^+dy^-)$, where $y^+\equiv x+iz$ and $y^-\equiv x-iz$. Hence the lattice AdS$_3$ spacetime interval can be constructed as: 
\bea
&&\big(\Delta s_{3\mathrm{l},j,k,l,m_1,m_2,m_3,m_4,m_5, m_6}\big)^2
\nn\\
&\equiv&\frac{4}{\Lambda }\Bigg(\frac{t_{j+m_1}-t_{j+m_4}}{f_{k+m_2,l+m_3}-f^*_{k+m_2,l+m_3}}
\nn\\
&&\times\frac{t_{j+m_1}-t_{j+m_4}}{f_{k+m_5,l+m_6}-f^*_{k+m_5,l+m_6}}
\nn\\
&&+\frac{f_{k+m_2,l+m_3}^*-f^*_{k+m_5,l+m_6}}{f_{k+m_2,l+m_3}-f^*_{k+m_2,l+m_3}}
\nn\\
&&\times\frac{f_{k+m_2,l+m_3}-f_{k+m_5,l+m_6}}{f_{k+m_5,l+m_6}-f^*_{k+m_5,l+m_6}}\Bigg)
\nn\\
&\equiv&g_{\mathrm{l}, tt,j,k,l,m_3,m_4,m_5,m_6}(t_{j+m_1}-t_{j+m_4})(t_{j+m_1}-t_{j+m_4})
\nn\\
&&+2g_{\mathrm{l},+-,j,k,l,m_3,m_4,m_5,m_6}
\nn\\
&&\times(f_{k+m_2,l+m_3}^*-f^*_{k+m_5,l+m_6})
\nn\\
&&\times(f_{k+m_2,l+m_3}-f_{k+m_5,l+m_6}),
\eea
where 
\bea
&&g_{\mathrm{l}, tt,j,k,l,m_3,m_4,m_5,m_6}
\nn\\
&\equiv&\frac{4}{\Lambda} 
\nn\\
&&\times\frac{1}{f_{k+m_2,l+m_3}-f^*_{k+m_2,l+m_3}}
\nn\\
&&\times \frac{1}{f_{k+m_5,l+m_6}-f^*_{k+m_5,l+m_6}},
\nn\\
&&g_{\mathrm{l}, +-,j,k,l,m_3,m_4,m_5,m_6}
\nn\\
&\equiv&\frac{2}{\Lambda}
\nn\\
&&\times\frac{1}{f_{k+m_2,l+m_3}-f^*_{k+m_2,l+m_3}}
\nn\\
&&\times\frac{1}{f_{k+m_5,l+m_6}-f^*_{k+m_5,l+m_6}},
\nn\\
&&f_{k,l}
\nn\\
&\equiv& x_k+iz_l,
\nn\\
&&x_{k\pm 1}
\nn\\
&\equiv& x_k\pm a_x,
\nn\\
&&z_{l\pm 1}
\nn\\
&\equiv& z_l\pm a_z, 
\nn\\
&&t_{j\pm 1}
\nn\\
&=&t_j\pm a_t.
\eea

\subsection{Lattice Embedding Coordinates}
We construct the lattice embedding coordinates in the lattice AdS$_2$ geometry and AdS$_3$ geometry to demonstrate the manifest isometry for each lattice spacing and size.

\subsubsection{Lattice AdS$_2$ Geometry}
The lattice embedding coordinates of the lattice AdS$_2$ metric are defined by that $-X_{1,j,k}^2-X_{2,j,k}^2+X_{3,j,k}^2\equiv 1/\Lambda$ for each index $j$ and $k$. We can show the lattice AdS$_2$ metric from these lattice embedding coordinates:
\bea
&&-(X_{1,j+m_1,k+m_2}-X_{1, j+m_3,k+m_4})^2
\nn\\
&&-(X_{2,j+m_1,k+m_2}-X_{2, j+m_3,k+m_4})^2
\nn\\
&&+(X_{3,j+m_1,k+m_2}-X_{3, j+m_3,k+m_4})^2
\nn\\
&=&\frac{4}{\Lambda }
\nn\\
&&\times\frac{g_{j+m_1,k+m_2}^*-g^*_{j+m_3,k+m_4}}{g_{j+m_1,k+m_2}-g^*_{j+m_1,k+m_2}}
\nn\\
&&\times\frac{g_{j+m_1,k+m_2}-g_{j+m_3,km_4}}{g_{j+m_3,k+m_4}-g^*_{j+m_3,k+m_4}}
\nn\\
&=&\big(\Delta s_{2\mathrm{l},j,k,m_1,m_2,m_3,m_4}\big)^2,
\eea
in which we used: 
\bea
X_{1,j,k}&\equiv& \frac{ig_{j,k}g_{j,k}^*}{g_{j,k}-g_{j,k}^*}-\frac{\frac{i}{\Lambda}}{g_{j,k}-g_{j,k}^*},
\nn\\
 X_{2,j,k}&\equiv&\frac{1}{\sqrt{|\Lambda|}}\frac{g_{j,k}+g_{j,k}^*}{g_{j,k}-g_{j,k}^*}, 
 \nn\\
 X_{3,j,k}&\equiv& \frac{ig_{j,k}g_{j,k}^*}{g_{j,k}-g_{j,k}^*}+\frac{\frac{i}{\Lambda}}{g_{j,k}-g_{j,k}^*}.
\eea
This shows that the lattice AdS$_2$ geometry manifestly preserves the isometry SO(2, 1) from the lattice embedding coordinates for each lattice spacing and size. 

\subsubsection{Lattice AdS$_3$ Geometry}
The lattice embedding coordinates of the lattice AdS$_3$ metric are defined by $-X_{1,j,k,l}^2-X_{2,j,k,l}^2+X_{3,j,k,l}^2+X_{4,j,k,l}^2\equiv 1/\Lambda$. The lattice AdS$_3$ metric can be rewritten in terms of these lattice embedding coordinates as the following:
\bea
&&-(X_{1,j+m_1,k+m_2,l+m_3}-X_{1, j+m_4,k+m_5,l+m_6})^2
\nn\\
&&-(X_{2,j+m_1,k+m_2,l+m_3}-X_{2, j+m_4,k+m_5,l+m_6})^2
\nn\\
&&+(X_{3,j+m_1,k+m_2,l+m_3}-X_{3, j+m_4,k+m_5,l+m_6})^2
\nn\\
&&+(X_{4,j+m_1,k+m_2,l+m_3}-X_{4, j+m_4,k+m_5,l+m_6})^2
\nn\\
&=&\frac{4}{\Lambda }\Bigg(\frac{t_{j+m_1}-t_{j+m_4}}{f_{k+m_2,l+m_3}-f^*_{k+m_2,l+m_3}}
\nn\\
&&\times\frac{t_{j+m_1}-t_{j+m_4}}{f_{k+m_5,l+m_6}-f^*_{k+m_5,l+m_6}}
\nn\\
&&+\frac{f_{k+m_2,l+m_3}^*-f^*_{k+m_5,l+m_6}}{f_{k+m_2,l+m_3}-f^*_{k+m_2,l+m_3}}
\nn\\
&&\times\frac{f_{k+m_2,l+m_3}-f_{k+m_5,l+m_6}}{f_{k+m_5,l+m_6}-f^*_{k+m_5,l+m_6}}\Bigg)
\nn\\
&=&\big(\Delta s_{3\mathrm{l},j,k,l,m_1,m_2,m_3,m_4,m_5,m_6}\big)^2,
\eea
in which we used: 
\bea
X_{1,j,k,l}&=&-\frac{2}{\sqrt{|\Lambda|}}\frac{t_j}{f_{k,l}-f_{k,l}^*},
\nn\\
X_{2,j,k,l}&=&-i\frac{\frac{1}{\Lambda}-t_j^2}{f_{k,l}-f_{k,l}^*}+i\frac{f_{k,l}f_{k,l}^*}{f_{k,l}-f_{k,l}^*},
\nn\\
X_{3, j,k,l}&=&-\frac{i}{\sqrt{|\Lambda|}}\frac{f_{k,l}+f_{k,l}^*}{f_{k,l}-f_{k,l}^*},
\nn\\
X_{4,j,k,l}&=&i\frac{f_{k,l}f_{k,l}^*}{f_{k,l}-f_{k,l}^*}+i\frac{\frac{1}{\Lambda}+t_{j}^2}{f_{k,l}-f_{k,l}^*}.
\eea
This lattice embedding coordinates also manifestly demonstrates the isometry SO(2,2) in the lattice AdS$_3$ geometry.

\subsection{Compactification}
We can also choose the lattice embedding coordinates: 
\bea
X_{1,j,k,l}&=&\frac{1}{\sqrt{|\Lambda|}}\cosh(\rho_{j,k,l})\cos(i\tau_{j,k,l}), 
\nn\\
X_{2,j,k,l}&=&\frac{1}{\sqrt{|\Lambda|}}\cosh(\rho_{j,k,l})\sin(i\tau_{j,k,l}), 
\nn\\
X_{3,j,k,l}&=&\frac{1}{\sqrt{|\Lambda|}}\sinh(\rho_{j,k,l})\sin(\theta_{j,k,l}),
\nn\\
X_{4,j,k,l}&=&\frac{1}{\sqrt{|\Lambda|}}\sinh(\rho_{j,k,l})\cos(\theta_{j,k,l}),
\eea
where $\rho_{j,k,l}>0$, $0\le \tau_{j,k,l}<2\pi$ and $0\le\theta_{j,k,l}<2\pi$, in the lattice AdS$_3$ metric. Therefore, we obtain the lattice AdS$_3$ spacetime interval
\bea
&&\big(\Delta s_{3\mathrm{l},j,k,l, m_1,m_2,m_3,m_4,m_5,m_6}\big)^2
\nn\\
&=&\frac{2}{\Lambda}
\nn\\
&&+\frac{2}{\Lambda}\bigg(-\cosh(\rho_{j+m_1,k+m_2,l+m_3})
\nn\\
&&\times
\cosh(\rho_{j+m_4,k+m_5,l+m_6})
\nn\\
&&\times\cos\big(i(\tau_{j+m_1,k+m_2,l+m_3}
-\tau_{j+m_4,k+m_5,l+m_6})\big)
\nn\\
&&+\sinh(\rho_{j+m_1,k+m_2,l+m_3})\sinh(\rho_{j+m_4,k+m_5,l+m_6})
\nn\\
&&\times\cos(\theta_{j+m_1,k+m_2,l+m_3}-
\theta_{j+m_4,k+m_5,l+m_6})\bigg).
\eea 
When we compactify the $\theta$ direction, we obtain the lattice spacetime interval \cite{Achucarro:1993fd}:
\bea
&&\big(\Delta s_{2\mathrm{l},j,k,l,m_1,m_2,m_3,m_4,m_5,m_6}\big)^2
\nn\\
&=&\frac{2}{\Lambda}
\nn\\
&&+\frac{2}{\Lambda}
\bigg(-\cosh(\rho_{j+m_1,k+m_2,l+m_3})
\nn\\
&&\times\cosh(\rho_{j+m_4,k+m_5,l+m_6})
\nn\\
&&\times\cos\big(i(\tau_{j+m_1,k+m_2,l+m_3}-\tau_{j+m_4,k+m_5,l+m_6})\big)
\nn\\
&&+\sinh(\rho_{j+m_1,k+m_2,l+m_3})\sinh(\rho_{j+m_4,k+m_5,l+m_6})\bigg)
\nn\\
\eea
and the lattice dilaton field
\bea
&&\phi_{j,k,l,m_1,m_2,m_3,m_4,m_5,m_6}
\nn\\
&=&\frac{1}{\sqrt{|\Lambda|}}
\nn\\
&&\times\sqrt{|\sinh(\rho_{j+m_1,k+m_2,l+m_3})\sinh(\rho_{j+m_4,k+m_5,l+m_6})|}.
\nn\\
\eea
The lattice spacetime interval $\big(\Delta s_{2\mathrm{l},j,k,l,m_1,m_2,m_3,m_4,m_5,m_6}\big)^2$ can also be rewritten in terms of the lattice embedding coordinates with the manifest isometry SO(2, 1) without suffering from the lattice artifact. Hence we obtain the lattice AdS$_2$ spacetime interval from the lattice AdS$_3$ spacetime interval through the compactification for each lattice spacing and size.
\\

Although we only wrote the construction of the lattice AdS$_2$ geometry and AdS$_3$ geometry, the generalization should be straightforward from the lattice embedding coordinates.

\section{Application to Tensor Network}
\label{3}
We first calculate the codimension two AdS$_3$ minimum surfaces at a given time slice from the perturbation with respect to the lattice spacing. This provides a well-defined perturbation limit around the minimum surface on a lattice, and then we use the minimum surface to discuss the holographic tensor network \cite{Swingle:2009bg}.

\subsection{AdS$_3$ Minimum Surface}
The lattice area $A_{\mathrm{AdS}_{3\mathrm{l}}}$ is given by
\bea
A_{\mathrm{AdS}_{3\mathrm{l}}}= \frac{1}{\sqrt{-\Lambda}}\sum_{j=1}^{n_A}\bigg(\frac{(f_{j+1}^*-f^*_{j-1})(f_{j+1}-f_{j-1})}{(f^*_{j+1}-f_{j+1})(f_{j-1}-f^*_{j-1})}\bigg)^{\frac{1}{2}},
\nn\\
\eea
where $n_A$ is a number of lattice points in the lattice area $A$. This lattice area is invariant under the isometry of AdS$_3$ metric for each finite lattice spacing and size.
\\

We solve the minimum surface equation $\delta A_{\mathrm{AdS}_{3\mathrm{l}}}/\delta z_j$ through the perturbation $z=z_0+a_x^2z_2+\cdots$. The solution $z(x)$ up to the second order of lattice spacing satisfies
$\big(z^{\prime\prime}/z+1/z^2+z^{\prime 2}/z^2\big)/(1+z^{\prime 2})^{\frac{3}{2}}
=2a_x^2z_0^{-7}Lx^2+\cdots$, where $-L< x< L$.
\\

The perturbative solution $z(x)$ can be solved exactly
\bea
&&z
\nn\\
&=&\sqrt{L^2-x^2}\bigg(1+a_x^2\frac{\frac{L^3(L^2+x^2)}{(L^2-x^2)^2}-3x\tanh^{-1}\big(\frac{x}{L}\big)-L}{24L(L^2-x^2)}\bigg)
\nn\\
&&+\cdots.
\eea
If one puts the cut-off $\delta$, which satisfies $\delta^2/L^2\ll 1$, to the interval as $x=L-\delta$, where $\delta$, one can find that
\bea
\frac{a_x^2}{12L}\frac{L^3(L^2+x^2)}{(L^2-x^2)^2}=\frac{a_x^2}{12}\frac{\big(1+\frac{x^2}{L^2}\big)}{\big(1-\frac{x^2}{L^2}\big)^2}=\frac{a_x^2}{\delta^2}\frac{L^2}{24}
+\cdots.
\nn\\
\eea
This should imply that the perturbation study should require the limit $a_x^2/\delta^2\ll 1$. If one choose $\epsilon\equiv z(L-\delta)$,
one should obtain $\epsilon=\sqrt{2\delta L}\bigg(1+a_x^2L/(96\delta^3)+\cdots\bigg)$.
Hence ones should require the other limit $a_x^2L/\delta^3\ll 1$ to do the perturbation to study the center charge \cite{Ryu:2006bv} on the lattice. These limits show that the cut-off on the interval $\delta$ cannot be at the same order of lattice spacing $a_x$ . If one only wants to do the perturbation to get the minimum surface, we only require the limits, $\delta^2/L^2\ll 1$ and $a_x^2/\delta^2\ll 1$. The remaining limit, $a_x^2L/\delta^3\ll 1$, helps us to obtain the lattice correction to the universal term \cite{Ryu:2006bv}. Using the lattice perturbation to construct the tensor network in AdS$_3$ background also requires the limit to obtain a consistent study.

\subsection{Holographic Tensor Network}
Without meeting the problem of continuum limit \cite{Bao:2015uaa}, we use the different identification between the lattice AdS$_3$ geometry and tensor network. Here we only consider the leading result but still follow the same limit as in the lattice perturbation.
\\

We first consider a horizontal line, $z_j=z_0$ for each $j$, in the AdS$_3$ metric. Therefore, the length of the horizontal line is:
\bea
|\gamma_{\mathrm{AdS}_31}|
=\frac{1}{\sqrt{-\Lambda}}\frac{2(L-\delta)}{z_0},
\eea
in which we define that the total length of the horizontal line is $2(L-\delta)$. Then we define that the the length of the horizontal line in the tensor network
\bea
&&|\gamma_{\mathrm{ts}1}|
\nn\\
&=&L_1\cdot(\mathrm{the\ number\ of\ bonds\ in\ the\ horizontal\ line}),
\nn\\
\eea
in which we assumed 
\bea
&&(\mathrm{the\ number\ of\ bonds\ in\ the\ horizontal\ line})
\nn\\
&\equiv&\frac{1}{a_x\sqrt{-\Lambda}}\frac{2(L-\delta)}{z_0},
\eea
$z_0=\epsilon$ is the cut-off of $z$, and $L_1$ is a horizontal proper length for each bond. We assume that $|\gamma_{\mathrm{AdS}_31}|=|\gamma_{\mathrm{ts}1}|$, and then we obtain $L_1=a_x$. If one considers the continuum limit $a_x\rightarrow 0$, one obtains $L_1\rightarrow 0$, and a number of bounds in the horizontal line should go to infinity.
\\

Now we discuss the renormalization group flow. The renormalization group flow begins from the boundary point $z_0=\epsilon$ to the bulk point, $z_m=k^m\epsilon$, where $k\equiv 1+\epsilon$, and $m$ is a positive integer, in the tensor network. Because we only study the leading order correction of lattice spacing $a_x$, we have $\delta=L-\sqrt{L^2-\epsilon^2}+\cdots$.
\\

Naively, the operation of renormalization group flow needs the number of steps
\bea
2\log_k\bigg(\frac{1}{a_x\sqrt{-\Lambda}}\frac{2(L-\delta)}{\epsilon}\bigg)
=2\log_k\bigg(\frac{1}{a_x\sqrt{-\Lambda}}\frac{2L}{\epsilon}\bigg)+\cdots.
\nn\\
\eea
Indeed, this number of operation is strange. When we only consider that the length of interval shrinks to zero, the number of operation should vanish, but the number of operation that we defined can always be infinity under the continuum limit $a_x\rightarrow 0$. Therefore, we consider the number of operation from the length $2L$ to the length $\epsilon$, $2\log_k(2L/\epsilon)$. We relate this number of operation to the number of bonds in the geodesic line
\bea
&&(\mathrm{the\ number\ of\ bonds\ in\ the\ geodesic\ line})
\nn\\
&\equiv& 2\log_k\bigg(\frac{2L}{\epsilon}\bigg).
\eea
The length of the geodesic line in the tensor network should be defined by:
\bea
|\gamma_{\mathrm{ts}2}|&\equiv& L_2\cdot(\mathrm{a\ number\ of\ bouds\ in\ the\ geodesic\ line})
\nn\\
&\equiv& 2L_2\log_k\bigg(\frac{2L}{\epsilon}\bigg),
\eea
where $L_2$ is a vertical proper length of the bonds in the tensor network. 
\\

 The length of a geodesic line in AdS$_3$ metric is
\bea
|\gamma_{\mathrm{AdS}_32}|=2\sqrt{-\frac{1}{\Lambda}}\ln\frac{2L}{\epsilon}+\cdots.
\eea
If we assume $|\gamma_{\mathrm{ts}2}|=|\gamma_{\mathrm{AdS}_32}|$, we obtain $L_2=(\ln k)/\sqrt{-\Lambda}$. When we take the limit $\epsilon\rightarrow 0$, we obtain $k\rightarrow 1$. Therefore, we get $L_2\rightarrow 0$. The above identification between the tensor network and lattice geometry provides the continuum limit to the tensor network.

\section{Outlook}
\label{4}
\noindent 
We constructed the lattice AdS geometry without losing the isometry. Because the lattice AdS$_2$ induced metric provided the lattice Schwarzian theory at the classical limit \cite{Chu:2018jje}, which has the continuum description, and the lattice AdS$_2$ geometry can be obtained from the higher dimensional lattice AdS geometry through the compactification without suffering from the lattice artifact, and the lattice AdS metric does not depend on the angular variables. Hence we conclude that the lattice Einstein gravity theory in the lattice AdS background should have the continuum limit at the IR limit. From the lattice embedding coordinates, it is easy to give the Lorentzian lattice AdS metric without restricting to a study of the Euclidean lattice AdS metric. We can only take the continuum limit in the time direction without the continuous space and apply the study to the Hamiltonian formulation on a lattice. It should be interesting to study the application of tensor network in the holographic renormalization group flow \cite{Swingle:2009bg} and the lattice correction in the holographic study \cite{tHooft:1993dmi}. The lattice construction also has generic applications because the construction is not restricted to the lattice AdS background, and the construction can be extended to other lattice backgrounds with an isometry.

\section*{Acknowledgments}
\noindent 
The author would like to thank Sinya Aoki, Su-Kuan Chu, Masanori Hanada, Nikita Nekrasov, Sarthak Parikh, and Tadashi Takayanagi for their useful discussions. 
The author was supported by the Post-Doctoral International Exchange Program and China Postdoctoral Science Foundation, Postdoctoral General Funding: Second Class (Grant No.
2019M652926) and would like to thank Nan-Peng Ma for his encouragement.
\\

\noindent 
The author thanks Tohoku University, Okinawa Institute of Science and Technology Graduate University, Yukawa Institute for Theoretical Physics at the Kyoto University, Istituto Nazionale Di Fisica Nucleare - Sezione di Napoli, Università degli Studi di Napoli Federico II, Kadanoff Center for Theoretical Physics at the University of Chicago, Stanford Institute for Theoretical Physics at the Stanford University, Kavli Institute for Theoretical Physics at the University of California of Santa Barbara, National Tsing Hua University, Israel Institute for Advanced Studies at the Hebrew University of Jerusalem, Jinan University, Institute of Physics at the University of Amsterdam, Shing-Tung Yao Center at the Southeast University, Institute of Theoretical Physics at the Chinese Academy of Sciences, Shanghai University, Shanghai Jiao Tong University, Sun Yat-Sen University, Institute for Advanced Study at the Tsinghua University, Center for Gravitation and Cosmology at the Yangzhou University, and Zhejiang Institute of Modern Physics at the Zhejiang University.
\\

\noindent 
Discussions during the workshops ``String-Math 2018'', ``Strings 2018'', ``New Frontiers in String Theory'', ``Strings and Fields 2018'', "Order from Chaos", "NCTS Annual Theory Meeting 2018: Particles, Cosmology and Strings", and "The 36th Jerusalem Winter School in Theoretical Physics - Recent Progress in Quantum Field / String Theory", ``Jinan University Gravitational Frontier Seminar'', ``Quantum Information and String Theory'', ``Strings 2019'', ``Amsterdam Summer Workshop on String Theory'', ``Youth Symposium on Theoretical High Energy Physics in Southeast University'', ``Workshop on Holography and Quantum Matter'', ``East Asia Joint Workshop on Fields and Strings 2019 and 12th Taiwan String Theory Workshop'', ``Composite 2019: Hunting Physics in Higgs, Dark Matter, Neutrinos, Composite Dynamics and Extra-Dimensions'', and ``NCTS Annual Theory Meeting 2019: Particles, Cosmology and Strings'', were useful to complete this work. 

\end{document}